\documentclass[prl,twocolumn,showpacs,letterpaper,showpacs,aps]{revtex4}
\usepackage[]{graphicx,amsmath,amssymb,amsfonts,latexsym,color,dcolumn,bm,ulem}
\usepackage{verbatim}
\usepackage[]{graphicx}

\begin{document}
\newcommand{\beq}{\begin{equation}}
\newcommand{\eeq}{  \end{equation}}
\newcommand{\bea}{\begin{eqnarray}}
\newcommand{\eea}{  \end{eqnarray}}
\newcommand{\bit}{\begin{itemize}}
\newcommand{\eit}{  \end{itemize}}
\newcommand{\jmax}{j_{\text{max}}}

\providecommand{\abs}[1]{\left\lvert#1\right\rvert}
\providecommand{\norm}[1]{\lVert #1 \rVert}
\providecommand{\moy}[1]{\langle #1 \rangle}
\providecommand{\bra}[1]{\langle #1 \rvert}
\providecommand{\ket}[1]{\lvert #1 \rangle}
\providecommand{\braket}[2]{\langle #1 \rvert #2 \rangle}

\title{Non-linear coupling between the two oscillation modes of a dc-SQUID}

\author{F. Lecocq$^1$, J. Claudon$^2$, O. Buisson$^1$ and P. Milman$^{3,4}$}

\affiliation{$^1$Institut N\'eel, C.N.R.S.- Universit\'e Joseph Fourier, BP 166, 38042 Grenoble-cedex 9, France}
\affiliation{$^2$CEA-CNRS-UJF joint group 'NanoPhysique et SemiConducteurs', CEA, INAC, SP2M, F-38054 Grenoble, France}
\affiliation{$^3$Univ. Paris-Sud 11, Institut de Sciences Mol\'eculaires d'Orsay (CNRS), B\^{a}timent 210--Campus d'Orsay, 91405 Orsay Cedex, France}
\affiliation{$^4$Laboratoire Mat\'eriaux et Ph\'enom\`enes Quantiques, CNRS UMR 7162, Universit\'e Paris Diderot, 75013, Paris, France}

\begin{abstract} 
We make a detailed theoretical description of the two-dimensional nature of a dc-SQUID, analyzing the coupling between its  two orthogonal phase oscillation modes. While it has been shown that the mode defined as ``longitudinal" can be initialized, manipulated and measured, so as to encode a quantum bit of information, the mode defined as  ``transverse" is usually repelled at high frequency and does not interfere in the dynamics.  We show that, using typical parameters of existing devices, the transverse mode energy can be made of the order of the longitudinal one. In this regime, we can observe  a strong coupling between these modes, described by an Hamiltonian providing a wide range of interesting effects, such as conditional quantum operations and entanglement.  This coupling also creates an atomic-like structure for the combined two mode states, with a V-like scheme.
\end{abstract}

\maketitle

The control of physical systems at the quantum level represents a significant technological challenge. But, this goal is of extreme interest since successful applications would allow us to probe fundamental aspects of quantum physics and to demonstrate the claimed advantages of quantum computation. A number of physical systems have shown, in the last decade, their potentiality as ``quantum devices". Examples are atoms in a high finesse cavity~\cite{HAROCHE}, trapped ions~\cite{Blatt_Nature2008} and solid state quantum circuits~\cite{Wendin_LTP2007,Korotkov_QIP2009,Hanson_RMP2007,Neumann_Science2008}. 

Superconducting circuits (SC) appear as suitable candidates to study controllable macroscopic quantum systems. During the last decade, many experiments on such devices have demonstrated effects previously observed on atomic and quantum optics experiments~\cite{Wendin_LTP2007,Korotkov_QIP2009}. Furthermore, the extreme flexibility in choosing the characteristic parameters of SC have made possible the realization of new interaction regimes, such as the ultra strong coupling regime~\cite{Peropadre_PRL2010,Niemczyk_Naturephys2010}. In these experiments, SC act as artificial atoms with {\it one} degree of freedom, described either as a 2-level system or a multilevel system. 

In this work, we show that a dc-SQUID with a large loop inductance is an artificial atom with {\it two} degrees of freedom. We make a theoretical study of the two orthogonal phase oscillation modes and analyse the coupling between them. One of the modes, called hereafter the longitudinal mode (LM), was extensively studied and demonstrated a large variety of quantum phenomena~\cite{Buisson_QIP2009}. For instance, using the two lowest energy states, all required operations necessary to define a qubit (coherent manipulation, initialization, coupling and measurement) have been performed~\cite{Hoskinson_PRL2009}. The quantum dynamics of the transverse mode (TM), which is perpendicular to LM, have never been studied theoretically or experimentally, although it presents appealing potential advantages, such as longer coherence times and an optimal point for energy relaxation. 

Moreover, we also show that the nature and the strength of the coupling terms between the two oscillation modes can be engineered to a large extent. They lead to a wealth of novel effects, that were initially demonstrated or proposed in atomic physics. Examples include quantum non-demolition measurements, the realization of conditional quantum operations, the controlled creation and measurement of entanglement between the two mode. In addition, the coupling term in the dc-SQUID can  result in a V-type 3-level structure, that has recently attracted a lot of interest~\cite{ReviewSCquantOptics}. All these potential experimental achievements involve realistic circuit parameters and should be observable with present day technology. 

\begin{figure}
\includegraphics[width=0.50\textwidth]{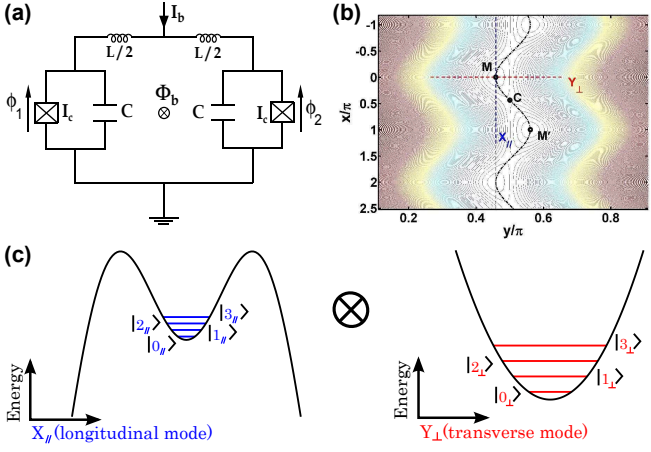}
\caption{(a) Electrical scheme of an inductive dc-SQUID (b) 2D potential for a dc-SQUID with $b=3$, $I_b=0$ and $\Phi_b=0.51\Phi_0$; $M$ and $M'$ indicate two different local minima separated by a saddle point $C$. The dashed black line represents the trajectory of minimum energy. (c) potential of the LM (left side) and TM (right side), related to the motion of the fictitious particle trapped at the local minimum $M$. \label{fig1}}
\end{figure}

A dc-SQUID is composed of two Josephson junctions (JJ) with critical current $I_0$, capacitance $C_0$ and phase difference $\phi_{1}$ and $\phi_{2}$, embedded in a loop of inductance $L$ (Fig.~\ref{fig1}a). The phase dynamics of this system can be mapped on the evolution of a fictitious particle of mass $m=2C_0\left (\Phi_0/2\pi \right )$, with two degrees of freedom $x=(\phi_1+\phi_2)/2$ and $y=(\phi_1-\phi_2)/2$, which is experiencing a potential \cite{Tesche_JLTP1977,Lefevre_PRB1992}:
\begin{eqnarray}
  \label{eq:1}
   U(x,y)  = U_0 \big[-\cos{x}\cos{y} -\frac{I_b}{2I_0} x+ b \Big( y-\frac{\pi\Phi_b}{\Phi_0} \Big)^2 \big],
\end{eqnarray}
where $\Phi_0=h/2e$ is the superconducting flux quantum, $U_0=2I_0\left (\Phi_0/2\pi \right )$, $I_b$ and $\Phi_b$ are respectively the current and flux bias and $b = \Phi_0/2\pi L I_0$ the ratio between the Josephson inductance and the loop inductance. For $I_b<2I_0$, this potential presents a periodic series of equivalent local minima separated by potential barriers (Fig.~\ref{fig1}b). For simplicity, we consider a perfectly symmetric dc-SQUID, with symmetric  inductances, JJs critical currents and capacitances.

To describe the 2D dynamics of the particle trapped in a potential minimum, we introduce the basis $(\mathbf{x}_{\parallel},\mathbf{x}_{\perp})$ corresponding to the motion parallel ($\parallel$) or transverse ($\perp$) to the trajectory of minimum energy~\cite{Fay_PRB2010}. It is obtained from a rotation with an angle $\theta(I_b,\Phi_b)$ of the basis $(\mathbf{x},\mathbf{y})$. The system dynamics is then governed by the Hamiltonian $\mathcal{H}=\mathcal{H}_{\parallel}+\mathcal{C}+\mathcal{H}_{\perp}$ which describes two anharmonic oscillators whose interaction is mediated by a coupling term $\mathcal{C}$ (Fig.~\ref{fig1}c). Up to order 4, $\mathcal{H}_{\alpha}= \hbar\omega_{\alpha} \times [ \tfrac{1}{2}( \hat p_{\alpha}^2+ \hat x_{\alpha}^2 ) -\sigma_{\alpha} \hat x_{\alpha}^3+\delta_{\alpha} \hat x_{\alpha}^4]$ with $\alpha = \parallel$ or $\perp$. Here, $\hat x_{\alpha}$ and $\hat p_{\alpha}$ are the reduced (dimensionless) position and momentum operators in the direction $\alpha$; they satisfy the commutation relation $[\hat x_{\alpha},\hat p_{\alpha}]=i\hat1$; the $\omega_{\alpha}$ are the characteristic frequencies of each mode; in general, they are different with $\omega_{\parallel} < \omega_{\perp}$. For relevant bias points, the oscillators are weakly anharmonic: $\delta_{\alpha} \ll 1$ and $\sigma_{\alpha}\ll 1$. Thus the energy $E_{n_{\alpha}}$ associated with the eigenstate $\ket{n_{\alpha}}$ of $\mathcal{H}_{\alpha}$ can be obtained from a simple perturbative approach. Specifically, $E_{n_{\alpha}}-E_{n_{\alpha}-1}=\left ( 1-\Lambda_{\alpha} n_{\alpha}\right) \hbar \omega_{\alpha}$ with $\Lambda_{\alpha}=\tfrac{15}{2}\sigma_{\alpha}+3 \delta_{\alpha}$ \cite{Hoskinson_PRL2009}. The coupling Hamiltonian reads $\mathcal{C}=\hbar g_{21}\hat x_{\parallel}^2 \hat x_{\perp}+\hbar g_{12} \hat x_{\parallel} \hat x_{\perp}^2+\hbar g_{22} \hat x_{\parallel}^2 \hat x_{\perp}^2$. Note that here the linear coupling term $\hat x_{\parallel} \hat x_{\perp}$ is cancelled by the $(\mathbf{x}_{\parallel},\mathbf{x}_{\perp})$ basis choice. The fourth order coupling terms $\hat x_{\parallel} \hat x_{\perp}^3$ and $\hat x_{\parallel}^3 \hat x_{\perp}$ are negligible at finite bias current and go to zero at zero bias current. We have also the frequency hierarchy $g_{21},g_{12},g_{22} \ll \omega_{\parallel},\omega_{\perp}$.

\begin{figure}
\includegraphics[width=0.49\textwidth]{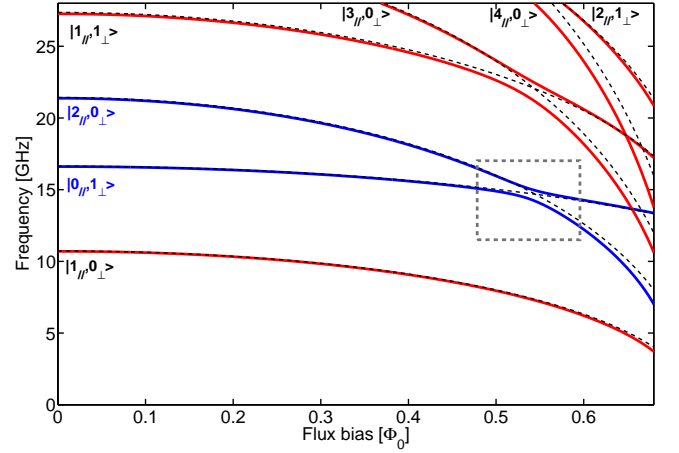}
\caption{Energy spectrum. Transition frequencies between ground state $\ket{0_{\parallel},0_{\perp}}$ and excited states $\ket{n_{\parallel},n_{\perp}}$ versus $\Phi_b$ at zero current bias of the artificial atom described by two oscillators without coupling (dashed lines) and taking into account the coupling (continuous lines) with $I_0=710\: $nA, $C_0=510\: $fF, $b=0.73$. The two solid blue curves highlight the states $\ket{{0_{\parallel}},1_{\perp}}$ and $\ket{{2_{\parallel}},0_{\perp}}$, chosen to illustrate the effect of the coupling term $\mathcal{H}_{\text{eff}}^{(2)}$. The area of maximum entanglement between these two states is found inside the dotted gray rectangle. \label{fig2}}
\end{figure} 

The LM alone was previously studied~\cite{Hoskinson_PRL2009,Claudon_PRB2008,Palomaki_PRB2010,Yu_PRB2010}. Quantum state detection was realized using a nanosecond pulse to produce a state dependent escape rate, through the potential barrier. At zero-current bias, the LM mode is insensitive to current noise~\cite{Hoskinson_PRL2009}. Nevertheless, away from the zero-flux bias point, it is still flux noise sensitive as other phase qubits~\cite{Lucero_PRL2008}. Until now, the TM has not been studied and the resonant dynamics of the dc-SQUID have been restricted to one dimension only. Indeed the plasma frequency of the TM  strongly increases with $b$. In previous dc-SQUID studies, the loop inductance was significantly smaller than the Josephson inductance, i.e $b>1$, leading to a high TM plasma frequency~\cite{Hoskinson_PRL2009,Palomaki_PRB2010}. By making the loop inductance larger than the Josephson inductance one can decrease the TM frequency and make it closer to the LM one, as shown in Fig.~\ref{fig2}. The TM presents several interesting theoretical features that demand to be experimentally investigated and exploited. For instance, current fluctuations induced by the environment lead to a TM relaxation rate $\Gamma_r=g_I(\theta)S_I(\omega_\perp)$ where $S_I(\omega_\perp)$ is the spectral density of the current noise and $g_I(\theta)$ the coupling strength between current and the TM. For a bias current $I_b=0$, this coupling goes to zero for the TM. Thus, a longer relaxation time is expected. Moreover, the dephasing time depends on $\frac{\partial \omega_\perp}{\partial I_b }$, which goes to zero at $I_b=0$, and on $\frac{\partial \omega_\perp}{\partial \Phi_b}$  which is one order of  magnitude smaller than for the LM (see Fig.~\ref{fig2}). Therefore, the TM is expected to suffer from a smaller dephasing than the LM.

In addition, the coupling between the TM and the LM leads to a number of physical effects, expanding the range of applicability of SC devices in quantum information field. Hereafter we will concentrate our study on these coupling effects. The Hamiltonian of the uncoupled degrees of freedom is given by $\mathcal{H}_0=\mathcal{H}_{\parallel}+\mathcal{H}_{\perp}$. The energies $E_{n_{\parallel},n_{\perp}} = E_{n_{\parallel}}+E_{n_{\perp}}$ of its eigenstates $\ket{n_{\parallel},n_{\perp}} = \ket{n_{\parallel}} \otimes \ket{n_{\perp}}$ are shown in Fig. 2 as dashed lines, for a particular set of circuit parameters. Since $g_{22},g_{12},g_{21} \ll \omega_{\perp},\omega_{\parallel}$, the interaction term $\mathcal{C}$ is treated as a perturbation. To perform a detailed analysis, we introduce the annihilation operators $\hat a$ and $\hat b$ of the LM and TM, and the occupation number operators $\hat n_{\parallel}=\hat a^{\dagger}\hat a$ and $\hat n_{\perp} = \hat b^{\dagger}\hat b$. One then obtains  $\mathcal{C}= \frac{1}{2\sqrt{2}}\hbar g_{21}(\hat a^2+\hat a^{\dagger 2}+2\hat n_{\parallel}+1)\times (\hat b +\hat b^{\dagger}) + \frac{1}{2\sqrt{2}}\hbar g_{12}(\hat b^2 +\hat b^{\dagger 2} + 2\hat n_{\perp}+1)\times(\hat a +\hat a^{\dagger})-\frac{1}{4} \hbar g_{22} (\hat a^2 + \hat a^{\dagger 2} + 2\hat n_{\parallel}+1 ) \times (\hat b^2 + \hat b^{\dagger 2}+2\hat n_{\perp}+1 )$. In the interaction picture, the interaction Hamiltonian is $\mathcal{H}_{\text{int}}= \mathcal{U} \mathcal{C} \mathcal{U}^{\dagger}$, with $\mathcal{U}=e^{-i(\mathcal{H}_{\parallel}+\mathcal{H}_{\perp})t}$. The effect of the coupling is captured by the matrix elements $\bra{{n'_{\parallel}},{n'_{\perp}}}\mathcal{H}_{\text{int}}\ket{{n_{\parallel}},{n_{\perp}}}=e^{-i \epsilon t}\bra{{n'_{\parallel}},{n'_{\perp}}}\mathcal{C}\ket{{n_{\parallel}},{n_{\perp}}}$, where  $\hbar \epsilon=E_{n'_{\parallel}}-E_{n_{\parallel}}+E_{n'_{\perp}}-E_{n_{\perp}}$.

In order to couple both modes in a non negligible way, $\bra{{n'_{\parallel}},{n'_{\perp}}}\mathcal{H}_{\text{int}}\ket{{n_{\parallel}},{n_{\perp}}}$  must oscillate at a frequency $\epsilon$ much slower than the coupling strength between the two states.
Only two terms can potentially verify this condition in $\mathcal{C}$: $\mathcal{H}_{\text{eff}}^{(1)} = \frac{1}{4}\hbar g_{22}\left (2\hat n_{\parallel}+1\right ) \times \left (2\hat n_{\perp}+1\right )$ and $\mathcal{H}_{\text{eff}}^{(2)} = \frac{1}{2\sqrt{2}}g_{21}(\hat b \hat a^{\dagger2} + \hat b^{\dagger}\hat a^2)$. The first term is present independently of the values of $n_{\parallel}$ and $n_{\perp}$. It induces a conditional energy shift for both modes. The second term leads to a single desexcitation of the TM and the simultaneous double excitation of the LM (and the hermitian conjugated process). It is relevant only when a quasi-resonant condition is verified between states $\ket{{n_{\parallel}},{n_{\perp}}}$ and $\ket{{n_{\parallel}\mp 2},{n_{\perp}\pm 1}}$. The deviation from the resonant condition between these two states is quantified by $\epsilon_{n_{\parallel}, n_{\perp}}(\Phi_b)=(E_{n_{\parallel}}+E_{n_{\perp}}-E_{n_{\parallel}-2}-E_{n_{\perp}+1})/\hbar \approx 2\omega_{\parallel}- \omega_{\perp} -(2n_{\parallel}+1)\Lambda_{\parallel}\omega_{\parallel}+n_{\perp}\Lambda_{\perp}\omega_{\perp}$. One can choose a set of circuit parameters so that $2\omega_{\parallel} \approx \omega_{\perp}$. As shown in Fig.~\ref{fig2}, tuning $\omega_{\parallel}$ and $\omega_{\perp}$ with the bias flux, we either satisfy the exact resonant condition ($\epsilon_{n_{\parallel},n_{\perp}}(\Phi_b)=0$), or ensure an off-resonant condition ($|\epsilon_{n_{\parallel},n_{\perp}}(\Phi_b)| \gg g_{21}$). We now study the {rich physics} associated with $\mathcal{H}_{\text{eff}}^{(1)}$ and $\mathcal{H}_{\text{eff}}^{(2)}$. Depending on the experimental parameters, each of these two terms can be chosen to dominate the coupling Hamiltonian.

We start with $\mathcal{H}_{\text{eff}}^{(1)}$, setting $|\epsilon_{n_{\parallel},n_{\perp}}(\Phi_b)| \gg g_{21}$, so that $\mathcal{H}_{\text{eff}}^{(2)}$ can be neglected. $\mathcal{H}_{\text{eff}}^{(1)}$ is very similar to the Kerr effect Hamiltonian~\cite{Butcher_1990}, with typical values of $ g_{22}/2\pi$ ranging from $50\: $MHz to $500\: $MHz. It allows i) the realization of quantum non-demolition (QND) measurement of the LM or TM \cite{QNDContinuous}, ii) conditional quantum logic (CQL) gates \cite{KerrQPG} and iii) the realization of a V-type 3-level atom, introduced initially in atomic-CQED. For example, QND measurement of the LM can be realized by TM spectroscopy: $\mathcal{H}_{\text{eff}}^{(1)}$ shifts in frequency the transition between the $\mathcal{H}_0$ eigenstates $\ket{ {n_{\parallel}}, 0_{\perp}}$ and $\ket{ {n_{\parallel}}, 1_{\perp}}$ proportionally to $(2n_{\parallel}+1)$. Using the two lowest energy states of both modes as qubits, this frequency shift can also be used to implement CQL gates. A microwave $\pi$-pulse resonant with the $\ket{{0_{\parallel}},1_{\perp}} \leftrightarrow \ket{{1_{\parallel}},1_{\perp}}$ longitudinal transition can switch both states without affecting $\ket{{0_{\parallel}},0_{\perp}}$ and $\ket{{1_{\parallel}},0_{\perp}}$. Therefore the qubit defined by the TM acts as a control qubit in a CNOT gate. Such gate have been successfully realized with flux qubit circuit~\cite{Plantenberg_2007}, but away from the optimal point. In the proposed dc-SQUID, such operations can be realized at the optimal point ($I_b=0$,$\Phi_b=0$), insensitive to flux and current noise in first order. Finally, the three states $\ket{{0_{\parallel}},0_{\perp}}$, $\ket{{1_{\parallel}},0_{\perp}}$ and $\ket{{0_{\parallel}},1_{\perp}}$ realize a V-like level structure, that was also proposed recently in charge qubit~\cite{Srinivasan_PRL2011}. As proposed in Ref.~\cite{ReviewSCquantOptics}, inserting such a V-type atom in a microwave resonator opens avenues for a wealth of effects discussed up to now in the context of atomic physics, such as quantum beats~\cite{BEATS}, the realization of dark states and the implementation of non-linear quantum field Hamiltonians~\cite{2A}.

\begin{figure}
\includegraphics[bb=7bp 13bp 300bp 220bp,clip,width=0.41\textwidth]{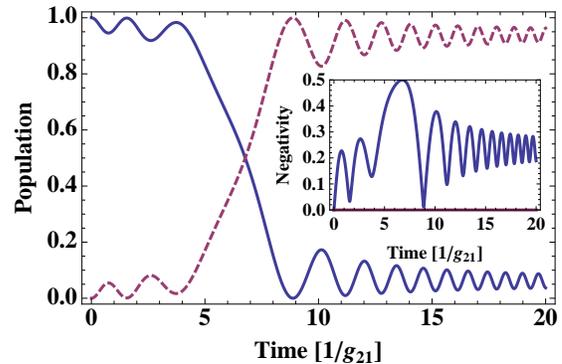}
\caption{(color online) Population transfer between states of the TM and the LM as a function of time for $\lambda=0.69 g_{21}^2$. Time is in units of $ g_{21}^{-1}$. The blue continuous line shows the population of state $ \ket{{0_{\parallel}},1_{\perp}}$, set equal to $1$ at $t=0$, and the pink dashed line is the population of state $ \ket{{2_{\parallel}},0_{\perp}}$, initially equal to $0$. The inset shows the absolute value of the negativity as a function of time, providing evidence that maximally entangled states between the two modes can be created. }
\label{fig3}
\end{figure}

We now discuss  $\mathcal{H}_{\text{eff}}^{(2)}$ in the case $ g_{22}\ll g_{21}$, so that  $\mathcal{H}_{\text{eff}}^{(1)}$ can be neglected. Experimentally, this condition is satisfied in a dc-SQUID with the parameters presented in Fig.~\ref{fig2}. In this case, $\mathcal{H}_{\text{eff}}^{(2)}$ couples states $\ket{{n_{\parallel}},n_{\perp}}$ to states $\ket{{n_{\parallel}\mp 2},n_{\perp}\pm 1}$ when $| \epsilon_{n_{\parallel},n_\perp}(\Phi_b)|\ll  g_{21}$. In the following, we show that $\mathcal{H}_{\text{eff}}^{(2)}$ can be used to i) create the Fock state $\ket{2_{\parallel}}$ in the longitudinal mode, starting from a single excitation quanta in the transverse mode, and  ii) entangle the states of the two oscillation modes.

To illustrate the first point, we consider the subspace spanned by the pair of states $\ket{{2_{\parallel}},{0_{\perp}}}$ and $\ket{{0_{\parallel}},1_{\perp}}$. The dependences of the associated energies on the bias flus are shown in Fig.~\ref{fig2} as solid blue curves. These two states satisfy the resonance condition at a bias flux $\Phi_r=0.54\Phi_0$, and $\mathcal{H}_{\text{eff}}^{(2)}$ then results in an avoided crossing (inside the gray rectangle). Away from the resonance, the coupling is negligible and the system eigenstates are those of the uncoupled Hamiltonian $\mathcal{H}_0$. Initially, $\Phi_b$ is set to $0$ and a $\pi$-pulse drives the TM from $\ket{{0_{\parallel}},0_{\perp}}$ to $\ket{{0_{\parallel}},1_{\perp}}$ (this state has a lower energy than $\ket{{2_{\parallel}},0_{\perp}}$). Increasing $\Phi_b$ slowly enough, it is then possible to realize an adiabatic passage through the avoided crossing. This results in the conversion of the state $\ket{{0_{\parallel}},1_{\perp}}$ into the state $\ket{{2_{\parallel}},{0_{\perp}}}$. To precise the adiabaticity condition, we suppose that in the vicinity of the resonance, the frequency difference between the states varies linearly in time: $\epsilon_{n_{\parallel},n_{\perp}}(\Phi_b) = \lambda t$. The adiabatic regime is obtained for a frequency sweep parameter $\lambda \ll (\pi/2)  g_{21}^2$.

This process is illustrated in Fig.~\ref{fig3} using realistic experimental parameters. The initial frequency splitting between the two involved states is $\epsilon_{2_{\parallel},0_{\perp}}(0)/2\pi \approx 5\: $GHz and $ g_{21}/2\pi \approx 600\: $MHz close to the level anti-crossing. With $\lambda=0.69 g_{21}^2$, corresponding to a variation of $1.8\: $GHz per nanosecond, one can perfectly transfer the population from $\ket{{0_{\parallel}},{1_\perp}}$ to $\ket{{2_{\parallel}},{0_\perp}}$. Experimentally, this frequency sweeping speed can be realized using a pulse generator with a typical rise-time of $5\: $ns. This time should be compared to the coherence damping time $T_1$. For a single mode dc-SQUID in the state $\ket{{1_{\parallel}}}$, $T_1\approx 100\: $ns~\cite{Hoskinson_PRL2009}. Since we expect the TM to present longer values of $T_1$, these relaxation times are more than enough to perform the state transfer protocol.

Here, we must stress that state transfer is made possible by the adiabatic evolution of $\Phi_b$. In the instantaneous limit ($\lambda \gg (\pi/2) g_{21}^2$), if the system is initially prepared in $\ket{{0_{\parallel}},1_{\perp}}$, for instance, it does not have the time to couple to state  $\ket{{2_{\parallel}},0_{\perp}}$ during the fast evolution. Consequently, coupling can be disregarded during the flux time variation. Such a condition is met using a pulse generator with a typical $500\: $ps rise-time.

Inspecting again Fig.~\ref{fig3} around the resonance, one observes strong oscillations in the population of $\ket{{0_{\parallel}},1_{\perp}}$. As we approach resonance, $\ket{{0_{\parallel}},1_{\perp}}$ is no longer an eigenstate of $\mathcal{H}_0$. Interestingly enough, the system eigenstates are in fact entangled states of the two oscillation modes. The inset of Fig.~\ref{fig3} shows the time evolution of entanglement, quantified by the negativity ${\cal N}$~\cite{NEGATIVITY}. During the adiabatic evolution, the system passes through a maximally entangled state, evidenced by ${\cal N}=1/2$. This consideration, together with the time evolution regimes explored in the previous paragraphs, suggest a procedure to create and detect controlled entanglement between the longitudinal and transverse modes. To prepare a maximally entangled state, for instance, one can stop the adiabatic evolution at the time corresponding to ${\cal N}=1/2$. By 'instantaneously' restoring the system to the non-resonant regime, it remains a maximally entangled state. One can then measure this state and its entanglement by measuring the population and coherences of the TM and the LM. The same type of analysis can be extended to other, more excited, TM-LM states, where coupling between modes can also occur depending on the value of $\epsilon_{n_{\parallel}, n_{\perp}}(\Phi_b)$.

The double excitation of one mode and the simultaneous single desexcitation of another mode (and the hermitian process), has been experimentally observed in cavity QED systems~\cite{2PHOTON}. In this system, the two modes  of a high quality microwave cavity  (formally equivalent to two harmonic oscillators) interact, mediated by an atom, via a third order process in perturbation theory. In trapped ion systems, such Hamiltonians can also be produced by tuning a laser to specific sideband frequencies corresponding to the excitation of the two-dimensional vibrational motion~\cite{IONS}. Usually, the coupling constant is proportional to the fourth power of the Lamb-Dicke parameter, so the coupling strength is significantly reduced. In the present system the non linear coupling is observed in the strong coupling limit, {\it i.e.} $g_{21} \gg 2\pi T_{1,2}^{-1}$. 

In conclusion, we have shown that a dc-SQUID is an artificial atom with two coupled degrees of freedom, the LM and TM. This coupling is fundamentally interesting and can be described by a number of effective Hamiltonians whose strengths vary according to experimentally tunable parameters. Each Hamiltonian gives rise to a number of applications, such as QND measurement, quantum logic, entanglement creation and the introduction of a more complex level structure to a superconducting artificial atom. Further applications arise when we consider coupling this system to a microwave cavity, opening the way to generalizing a larger collection of results already existing in cavity QED.

We thank W. Guichard and F. W. J. Hekking for fruitful discussions.  This work was supported by the european EuroSQIP and SOLID projects and by the french ANR \textquotedblright QUANTJO\textquotedblright.


\begin{thebibliography}{0}
\expandafter\ifx\csname natexlab\endcsname\relax\def\natexlab#1{#1}\fi
\expandafter\ifx\csname bibnamefont\endcsname\relax
  \def\bibnamefont#1{#1}\fi
\expandafter\ifx\csname bibfnamefont\endcsname\relax
  \def\bibfnamefont#1{#1}\fi
\expandafter\ifx\csname citenamefont\endcsname\relax
  \def\citenamefont#1{#1}\fi
\expandafter\ifx\csname url\endcsname\relax
  \def\url#1{\texttt{#1}}\fi
\expandafter\ifx\csname urlprefix\endcsname\relax\def\urlprefix{URL }\fi
\providecommand{\bibinfo}[2]{#2}
\providecommand{\eprint}[2][]{\url{#2}}

\end{thebibliography}


\begin{thebibliography}{99}

\bibitem{HAROCHE} J.M. Raimond, M. Brune and  S. Haroche, Rev. Mod. Phys. {\bf  73},  565 (2001). 

\bibitem{Blatt_Nature2008} R. Blatt and D. Wineland, Nature {\bf 453}, 1008 (2008).

\bibitem{Wendin_LTP2007} G. Wendin and V. S. Shumeiko,  Low Temp. Phys. {\bf  33}, 724 (2007)

\bibitem{Korotkov_QIP2009} A. Korotkov, Quantum Inf. Process, {\bf  8}, 51 (2009).

\bibitem{Hanson_RMP2007} R. Hanson {\it et al.}, Rev. Mod. Phys {\bf  79}, 1217 (2007).

\bibitem{Neumann_Science2008} P. Neumann {\it et al.}, Science {\bf  320}, 1326-1329 (2008).

\bibitem{Peropadre_PRL2010} B. Peropadre {\it et al.}, Phys. Rev. Lett. {\bf 105}, 023601(2010).

\bibitem{Niemczyk_Naturephys2010} T. Niemczyk {\it et al.}, Nature Physics, {\bf  6}, 772 (2010).

\bibitem{Buisson_QIP2009} O. Buisson {\it et al.}, Quantum Inf. Process, {\bf  8}, 155 (2009).

\bibitem{Hoskinson_PRL2009} E. Hoskinson {\it et al.}, Phys. Rev. Lett. {\bf 102}, 097004(2009).

\bibitem{ReviewSCquantOptics} J. Q. You and F. Nori, Nature, {\bf 474}, 589 (2011).

\bibitem{Tesche_JLTP1977} C. D. Tesche and J. Clarke, J. Low Temp. Phys. {\bf 29}, 301 (1977).

\bibitem{Lefevre_PRB1992} V. Lefevre-Seguin, et al, Phys. Rev. B {\bf 46}, 5507 (1992).

\bibitem{Fay_PRB2010} A. Fay {\it et al.}, Phys. Rev. B {\bf 83}, 184510 (2011).

\bibitem{Palomaki_PRB2010} T.A. Palomaki, et al, Phys. Rev. B {\bf 81}, 144503 (2010).

\bibitem{Yu_PRB2010} H. F. Yu, et al, Phys. Rev. B {\bf 81}, 144518 (2010).

\bibitem{Claudon_PRB2008} J. Claudon {\it et al.},  Phys. Rev. B {\bf 78}, 184503 (2008).

\bibitem{Lucero_PRL2008} E. Lucero {\it et al.}, Phys. Rev. Lett. {\bf 100}, 247001(2008).

\bibitem{Butcher_1990} P. N. Butcher and D. N. Cotter, The Elements of Non-linear Optics (Cambridge University Press, Cambridge,1990).


\bibitem{QNDContinuous} N. Imoto, H. A. Haus and Y. Yamamoto, Phys. Rev. A {\bf 32}, 2287 (1985). 

\bibitem{KerrQPG} P. Kok, Phys. Rev. A {\bf 77}, 013808 (2008).

\bibitem{Plantenberg_2007} J. H. Plantenberg {\it et al.}, Nature, {\bf 447}, 836 (2007).

\bibitem{NEGATIVITY} G. Vidal and R. F. Werner, Phys. Rev. A {\bf 65}, 032314 (2002).

\bibitem{2PHOTON}  P. Bertet {\it et al.}, Phys. Rev. Lett. {\bf 88}, 143601 (2002).

\bibitem{IONS} R. L. de Matos Filho and W. Vogel, Phys. Rev. A {\bf 58}, R1661 (1998). 

\bibitem{Srinivasan_PRL2011} S. J. Srinivasan {\it et al.}, Phys. Rev. Lett. {\bf 106}, 083601 (2011).

\bibitem{BEATS} M. O. Scully and S.-Y. Zhu and A. Gavrielides, Phys. Rev. Lett. {\bf 62}, 2813 (1989). 

\bibitem{2A} L. Davidovich {\it et al.}, Phys. Rev. A {\bf 36}, 3771 (1987).





\end{thebibliography}
\end{document}